\documentclass{llncs}
\usepackage[sectionbib,authoryear]{natbib}
\usepackage{graphicx}
\usepackage{multicol}
\usepackage{color}
\usepackage{amsmath}
\usepackage{hyperref}

\begin{document}
\title{Improving Nanopore Reads Raw Signal Alignment}
\author{Vladim\'\i{}r Bo\v{z}a
\and Bro\v{n}a Brejov\'a \and Tom\'a\v{s} Vina\v{r}}
\institute{Faculty of Mathematics, Physics, and Informatics,
           Comenius University,\\
           Mlynsk\'a dolina, 842 48 Bratislava, Slovakia\\
           \{boza,brejova,vinar\}@fmph.uniba.sk 
}
\maketitle

\begin{abstract}
We investigate usage of dynamic time warping (DTW) algorithm for aligning
raw signal data from MinION sequencer. DTW is mostly using for fast alignment
for selective sequencing to quickly determine whether a read comes from sequence
of interest.

We show that standard usage of DTW has low discriminative power mainly due
to problem with accurate estimation of scaling parameters.
We propose a simple variation of DTW algorithm, which does not suffer from scaling
problems and has much higher discriminative power.
\end{abstract}

\section{Introduction}

In this paper, we propose improvements to algorithms for aligning
raw signals from MinION nanopore sequencer.
The MinION device by Oxford Nanopore \citep{minion1},
weighing only 90 grams, is
currently the smallest high-throughput DNA sequencer. 
Thanks to its low capital costs, small size and the possibility 
of analyzing the data in real time as they are produced, MinION is very
promising for clinical applications, such as monitoring infectious
disease outbreaks \citep{Quick2015,Quick2016},
and characterizing structural variants in cancer \citep{Norris2016}.

One of the MinION advantages is 
selective sequencing, where we only sequence "interesting" reads by
rejecting all other reads after reading the first few hundred bases.
The idea was mostly explored in Readuntil tool \citep{readuntil}.

Hard part of selective sequencing is deciding which reads to reject.
Standard algorithm for selective sequencing would be to base call the first few hundreds of events from the
read and then align them to the reference sequence. After aligning, we would accept the read if
we find a match with the acceptable similarity. 

Unfortunately, the speed of the base calling algorithm on a reasonable computer
is much lower than the rate at which the sequencer produces the data.
This means that we need to use other approaches than base calling followed by alignment.
We will explore an approach which does not translate electric signal
into DNA, but instead it works directly with the electric signal data.

\citet{readuntil} solved this problem using dynamic time warping
algorithm (DTW), but he only align reads to small target sequences with size
up to hundred kilobases.

When using DTW to align raw signal to larger targets we found out that
it does not have enough discriminative power to distinguish false matches from true
matches.
We hypothesize that the main problem lies in determining correct scaling and shift of the data (which is hard
on small number of events). We test this experimentaly by using scaling and shift
parameters determined from the whole read and our experiments agree with the hypothesis.
We propose a variant of DTW algorithm and demonstrate that it has much better
discriminative power than the original DTW algorithm.

\section{Dynamic Time Warping}

Using emission distributions from the Nanocall \citep{David2016}, we can construct
the expected electric signal for given sequence of DNA bases. 
Thus, our main goal is to compare two sequences of electric signals, characterized by event means.
We will ignore event standard deviation in this section.
More formally,
given an expected sequence
$e_1, e_2, \dots, e_n$ and an observed sequence $o_1, o_2, \dots, o_m$, we
need to determine whether the observed sequence fits into the expected sequence,
and if so, find the alignment.

Since the event detection is unreliable, the expected and observed sequences may not only exhibit small differences
in signal levels, but also some events may be missing or duplicated. 
A standard algorithm for comparing two such signal sequences, which vary in speed is called
Dynamic Time Warping (DTW) \citep{dtw}. DTW was first used for aligning speeches. 

DTW calculates a matching between elements of two sequences.
It produces a sequence of pairs $(a_1, b_1), \dots, (a_k, b_k)$, where $a_i$ represents an index
in the first signal sequence and $b_i$ represents an index in the second signal sequence.
The following properties should hold:
\begin{itemize}
\item $b_1 = 1$, $b_k = m$
\item $0 \leq b_i - b_{i-1} \leq 1$ 
\item $0 \leq a_i - a_{i-1} \leq 1$
\item $(a_i > a_{i-1}) \lor (b_i > b_{i-1})$ 
\end{itemize}


The goal of DTW is to minimize the cost of the matching, which
is the sum of costs for each matched pair.
The cost of matching a pair is defined by cost function $d(x,y)$ which compares two matched elements
from the two sequences. In our case, we have chosen simply $d(x,y) = (x-y)^2$.

The optimal matching can be found by using a simple dynamic programming.
We define subproblem 
$D[i][j]$ as the best cost of DTW for sequences $e_1, \dots, e_i$ and $o_1, \dots, o_j$, assuming that $e_i$ and $o_j$ are matched
together.
Using this subproblem, the dynamic programming would proceed according to the 
following recurrence:
$$D[i][j] = d(e_i, o_j) + \min(D[i-1][j-1], D[i-1][j], D[i][j-1])$$

In our application, we are checking whether the second sequence can be found in the first sequence,
thus the cost of the optimal alignment can be found as: $\min_i D[i][m]$.

The algorithm has $O(nm)$ time complexity, where $n$, $m$ are the lengths of the first and second
sequence.

\subsection{Applying DTW to Selective Sequencing}

Before running DTW on MinION data, we first have to determine correct scaling and shift of the
data, since it is different for each read. Standard procedure for this is Z-score normalization, i.e. normalizing data mean to zero and
variance to one, as used in Readuntil \citep{readuntil}.

After finding the best matching, we have to decide based on the score, 
whether the read matches the target sequence or whether the score only represents a spurious match of the query
to the target sequence (a false positive).
This is usually done by simply thresholding on the score.
We will call this setup a \emph{naive scaling}.

In the following sections, we demostrate that the naive scaling has very low
specificity when higher sensitivity is required.
This causes problems especially when we are
trying to match data to a longer reference sequence, where many false positive matches can occur.


Here, we propose a simple heuristics, which greatly improves the tradeoff between
sensitivity and specificity.
If the alignment of the query was fixed, we could easily 
find better scaling parameters by looking at the sequence of matched pairs 
and setting scaling and shift to minimize the matching cost.
More formally, given a DTW alignment $(a_1, b_1), \dots, (a_k, b_k)$, we are looking for parameters
$A, B$ such that the matching cost
$\sum_k (Ao_{b_i} + B - e_{a_i})^2$
is as small as possible. This can be solved by the
ordinary least squares regression.


Once we have better scaling parameters, it is possible to improve the matching
by running the DTW algorithm on newly rescaled signal. In fact, we could run several
iterations of rescaling and DTW to improve the matching and the scaling parameters.
We call this approach \emph{iterative scaling}.

In the next section, we compare the performance of the naive scaling and our new iterative
rescaling method.

\section{Experiments}

\begin{table}
\centering
\begin{tabular}{c|c|c}
Algorithm variant & Sensitivity & Specificity \\\hline
Baseline    & $90\%$ & $99.72\%$ \\
Naive           & $90\%$ & $99.88\%$ \\
Simple rescaling       & $90\%$ & $\mathbf{99.94\%}$ \\
Two iteration rescaling & $90\%$ & $99.92\%$ \\
\hline
Baseline    & $95\%$ & $99.64\%$ \\
Naive           & $95\%$ & $99.45\%$ \\
Simple rescaling       & $95\%$ & $\mathbf{99.88\%}$ \\
Two iteration rescaling & $95\%$ & $99.86\%$ \\
\hline
Baseline    & $98\%$ & $99.46\%$ \\
Naive           & $98\%$ & $90.74\%$ \\
Simple rescaling       & $98\%$ & $\mathbf{99.58\%}$ \\
Two iteration rescaling & $98\%$ & $99.46\%$ \\
\hline
Baseline    & $99\%$ & $98.88\%$ \\
Naive           & $99\%$ & $47.30\%$ \\
Simple rescaling       & $99\%$ & $95.88\%$ \\
Two iteration rescaling & $99\%$ & $\mathbf{97.20\%}$ \\
\end{tabular}
\caption{Comparison of specificity of several scaling variants on prespecified sensitivity levels.}
\label{table:dtwexp}
\end{table} 

In our experiments we have used R9 Ecoli dataset from Loman Labs \citep{ecolidata9} using
only reads with and alignment to the reference and longer than 3000 bases.

As a query, we have used events number 100 to 350 from each read.
In each experiment, we run each method for each read against aligned location and against hundred random 500 bp
long locations from the genome.
After running each method, we study specificity-sensitivity tradeoff of each method by varying the threshold
for the score. For each method, we set the threshold to achieve a predefined sensitivity (how many
times DTW accepted a match of the read to
the correct location) and measure specificity (how many times DTW rejected a match of the read to a random
location).

\begin{figure}
\begin{center}
\includegraphics[width=0.49\textwidth,trim=0.38cm 0 0.5cm 0, clip]{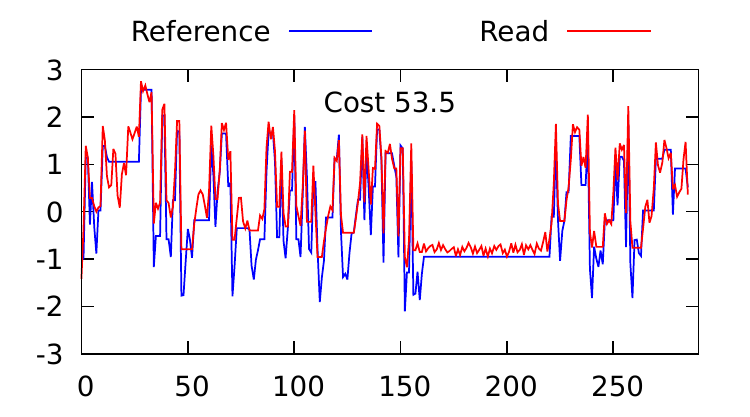}
\includegraphics[width=0.49\textwidth,trim=0.2cm 0 0.5cm 0, clip]{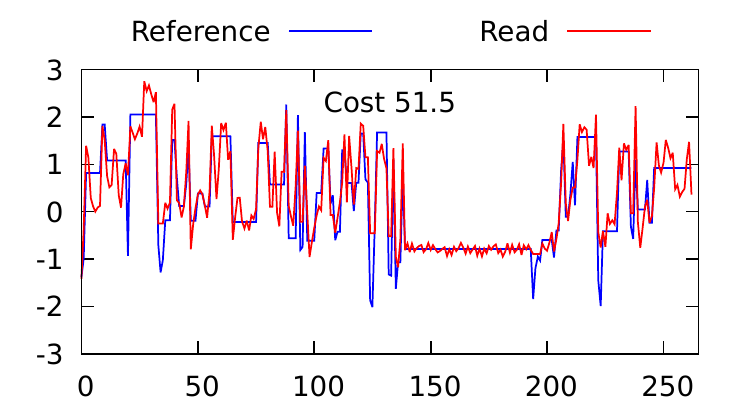}

\includegraphics[width=0.49\textwidth,trim=0.38cm 0 0.5cm 0, clip]{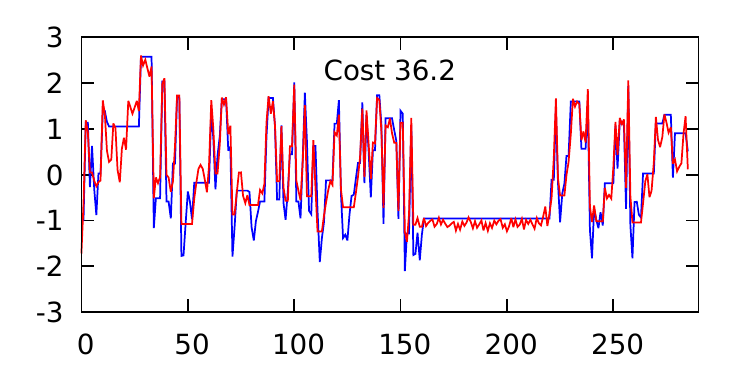}
\includegraphics[width=0.49\textwidth,trim=0.2cm 0 0.5cm 0, clip]{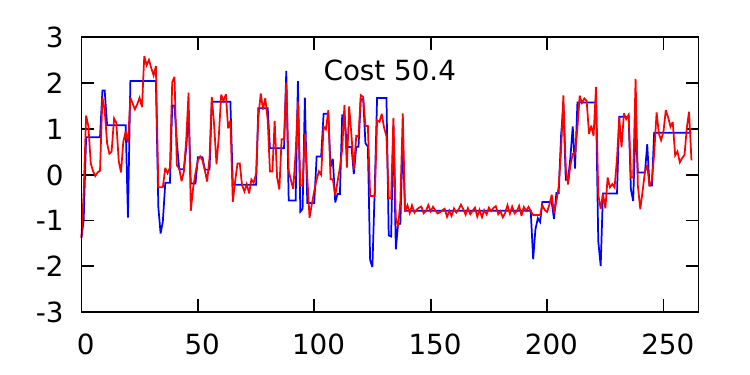}

\includegraphics[width=0.49\textwidth,trim=0.38cm 0 0.5cm 0, clip]{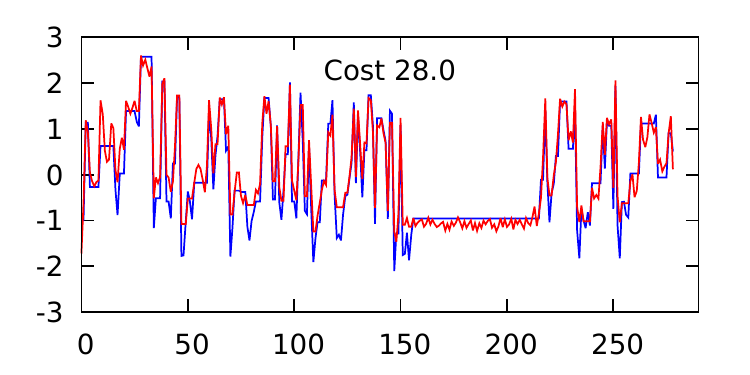}
\includegraphics[width=0.49\textwidth,trim=0.2cm 0 0.5cm 0, clip]{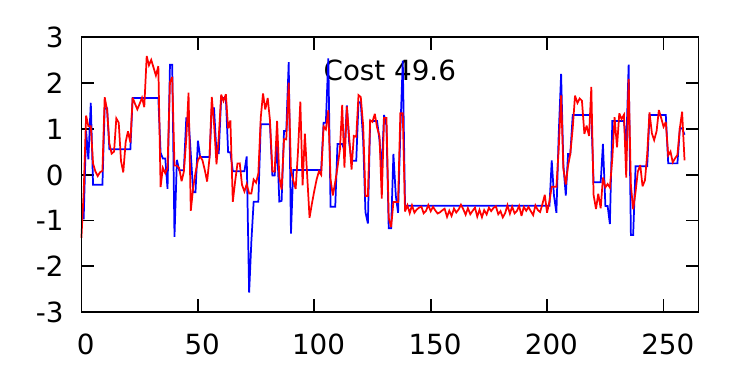}
\end{center}

\caption{Signal alignment after first DTW run (top), rescaling (middle), second DTW run (bottom).
True positive in the left column, false positive match in the right column.}
\label{fig:align}
\end{figure}

We compared the following methods:
\begin{itemize}
\item Z-score normalization using complete reads (on more than 3000 events), followed by DTW (on 250
events). This is an
unrealistic scenario in which we effectively know the correct scaling parameters. We will use this method as a \emph{baseline} for comparison.
\item Naive scaling, which represents the state of the art, consisting of Z-score normalization, followed by DTW.
\item Simple rescaling, consisting of Z-score normalization, followed by DTW and rescaling. 
\item Two iteration rescaling, consisting of Z-score normalization, followed by DTW, rescaling, and another DTW.
\end{itemize}

Figure \ref{fig:align} shows signal alignment progression of two iteration rescaling for true and false match.
Table \ref{table:dtwexp} shows comparison of these scaling variants.


We can see that at higher sensitivities, the naive variant performs much worse than rescaled variants
of DTW. This is especially visible at $99\%$ sensitivity level, where naive variant treats every second false match as a good match.
Also at $99\%$ sensitivity we see benefit of two iteration rescaling, which discards many more false positives than simple rescaling.
A consistently strong baseline performance suggests that the scaling is a big problem when using DTW
and need to be addressed before the method can be widely applicable. Our approach seems like a step
in a good direction to solve the problem.

\section{Conclusion and Further Research}

We have presented and validated improvements of signal
alignment for raw Nanopore sequencing data.
Our method is conceptually very simple and greatly improves
discriminative power of the alignment.

The obvious direction of further research is to develop
a data structure for indexing raw signals, where we could first preprocess target reference sequence
and then find candidate positions for possible signal matches for a query sequence, using
this index similarly as in classical sequence aligners \citep{bwamem}.

Also, an interesting topic to study theoretically is a variation of DTW, where we allow
the algorithm to select best scaling and shift parameters.

\paragraph{Acknowledgements.} This research was funded by
VEGA grants 1/0684/16 (BB) and 1/0719/14 (TV), and a grant
from the Slovak Research and Development Agency APVV-14-0253.
The authors thank Jozef Nosek for involving us in the MinION Access Programme,
which inspired this work.

\bibliographystyle{apalike} \bibliography{main}

\end{document}